# Thickness-dependent in-plane polarization and structural phase transition in van der Waals Ferroelectric CuInP$_2$S$_6$

*Jianming Deng*[1,#], *Yanyu Liu*[1,#], *Mingqiang Li*[2,#], *Sheng Xu*[3], *Yingzhuo Lun*[1], *Peng Lv*[1], *Tianlong Xia*[3], *Peng Gao*[2,4,\*], *Xueyun Wang*[1,\*], *Jiawang Hong*[1,\*]

[1]School of Aerospace Engineering, Beijing Institute of Technology, Beijing, 100081, China
[2]Electron microscopy laboratory, and International Center for Quantum Materials, School of Physics, Peking University, Beijing 100871, China
[3]Department of Physics and Beijing Key Laboratory of Opto-electronic Functional Materials & Micro-nano Devices, Renmin University of China, Beijing 100871, China
[4]Collaborative Innovation Centre of Quantum Matter, Beijing 100871, China

E-mail: p-gao@pku.edu.cn, xueyun@bit.edu.cn, hongjw@bit.edu.cn

**ABSTRACT**

Van der Waals (vdW) layered materials have rather weaker interlayer bonding than the intra-layer bonding, therefore the exfoliation along the stacking direction enables the achievement of monolayer or few layers vdW materials with emerging novel physical properties and functionalities. The ferroelectricity in vdW materials recently attracts renewed interest for the potential use in high-density storage devices. As the thickness going thinner, the competition between the surface energy, depolarization field and interfacial chemical bonds may give rise to the modification of ferroelectricity and crystalline structure, which has limited investigations. In this work, combining the piezoresponse force microscope scanning, contact resonance imaging, we report the existence of the intrinsic in-plane polarization in vdW ferroelectrics CuInP$_2$S$_6$ (CIPS) single crystals, whereas below a critical thickness between 90-100 nm, the in-plane polarization disappears. The Young's modulus also shows an abrupt stiffness at the critical thickness. Based on the density functional theory calculations, we ascribe these behaviors to a structural phase transition from monoclinic to trigonal structure, which is further verified by transmission electron microscope technique. Taken together, these findings demonstrate the foundational importance of structural phase transition for enhancing the rich functionality and broad utility of vdW ferroelectrics.

**1. Introduction**

One of the most common ways to achineve the mono-/few- layer van der Waals (vdW) is through mechnical exfolaition, which significantly reduces the thickness of specimens. Usually the phyiscal properties of vdW materials are thickness dependent,[1] and are also intriguing due to the scientific importance and minimization of potential devices applications. Take $MoTe_2$ as an illustration, when this system was exfoliated from bulk to few layered, a semimetal to semiconductor phase transition has been observed, which reveals the criticalness of Peierls distortion in the structural stabilization and unlocks the possible utilization in topological quantum devices.[2] More importantly, numerous novel phenomena and interesting applications can be anticipated if two-dimensional (2D) vdW materials undergo phase transitions in which ferroic functionalities, such as long-range magnetic[3-5] or ferroelectric ordering[6,7] possess various features in different phases. Regarding the low temperature of the 2D magnetic ordering, ferroelectric property opens the potential pathway for high-density storage devices at room temperature.[8] As the size reaching below the critical correlation length, due to the variation of surface energy, electrostatic doping, *etc.*, there exists possible structural phase transition in ferroelectric vdW materials with modification of ferroelectric polarization. Such manipulation and related functionalities are imperative, but not yet fully understood.

Besides the chemical functionalization of prevalent 2D materials into ferroelectric state,[9,10] 2D vdW materials with intrinsic ferroelectricity are more critical,[11,12] which may shed lights on the underlying physics of 2D ferroelectricity. 2D materials with in-plane ferroelectric polarization are more favored, such as ultrathin SnTe films,[13,14] hybrid perovskite

bis(benzylammonium) lead tetrachloride,[15] since the IP ferroelectricity avoid the effect from the increase of the depolarization field, which usually believed to be the critical reason for the disappearance of out-of-plane (OP) ferroelectricity at 2D limit. However, systems with OP ferroelectricity are also discovered[16-18] and more preferred in the potential devices applications. Meanwhile, the utilization of 2D OP ferroelectricity in the memory-devices requires the understanding of ferroelectric domains, switching behavior in the presence of external electric fields and the stability of reversed ferroelectricity. So far there are very few systems, such as $CuInP_2S_6$ (CIPS),[7,16,19-21] $In_2Se_3$,[6,8,17,18] $WTe_2$[22] and $MoTe_2$,[23] are discovered to host layered-perpendicular polarization.

In this work, we chose CIPS as a standard model to investigate the structural phase transition, as well as the evolution of ferroelectricity on the dependence of the thickness. CIPS belongs to a family of metal thio(seleno)phosphates, in which the variability in their structure leads to plenty of crystalline symmetries, which may modify the orientation of ferroelectricity.[24] Moreover, it is widely recognized that CIPS only has ferroelectric polarization perpendicular to the layered plane.[7,16,19,20] In this work, surprisingly, we found there also exists in-plane (IP) polarization in CIPS, and more interestingly, the IP polarization disappears below a critical thickness between 90-100 nm. The elastic property also shows abrupt stiffness change below the critical thickness. According to the density functional theory (DFT) calculations, a structural phase transition of CIPS from space group 7 (*Cc*) to 159 (*P*31*c*) was proposed below the critical thickness, which was verified from the scanning transmission electron microscopy (STEM) experiment. The coexist of IP and OP polarizations and new structure phase of CIPS below critical thickness may shed light on the designing and optimizing the piezoelectricity in layered materials.

## 2. Results and discussion

The single-crystals of CIPS were grown by chemical vapor transport method (see Experimental Section). The crystals formed in green transparent thin plates with lateral dimensional up to 1 cm, as shown in the inset of **Figure 1**a. X-ray diffraction pattern on the surface of single crystal indicates the *ab*-plane of CIPS (Figure 1a). In order to confirm the homogeneity and the composition of CIPS, STEM energy-dispersive X-ray spectroscopy (EDS) analysis were carried out, as shown in Supporting Information **Figure S1**. The elemental mapping images confirmed that Cu, In, P, and S are uniformly distributed with no noticeable compositional variation.

We performed piezoresponse force microscope (PFM) measurements (see Experimental Section) to investigate the ferroelectricity of bulk CIPS crystals. Switchable ferroelectric polarization with well-defined butterfly loops of the saturated PFM amplitude and the distinct 180° switching of the phase signals are observed, as shown in Figure 1b. The offset of the loop from zero bias originates from the Schottky barrier difference between the upper (CIPS and Ti/Ir conductive tip) and lower (CIPS/electrodes) interfaces of the sample. We observed the randomly distributed ferroelectric domains as Supporting Information **Figure S2** shows. Surprisingly, we not only detected the previously reported OP ferroelectric domains in CIPS bulk crystal,[16,19] but also we found there co-exists IP ferroelectric polarizations which has not been reported in this system before (Supporting Information Figure S2). In order to verify the IP ferroelectric domain, we performed domain switching PFM on mechanically exfoliated CIPS with various thicknesses, which were transferred onto heavily doped silicon substrates. We selected 320 and 90 nm as examples to show the domain switching behaviors (see Supporting Information **Figure S3** for 400 nm). The OP and IP phase images of 320 nm thick CIPS flake were acquired after writing box-in-box with opposite tip voltages (-8 and +10 V). Clear reversal of phase contrast demonstrates polarization direction can be controlled by the external bias, as shown in Figure 1c and 1d. More importantly, the IP phase changes simultaneously with the OP phase, indicating that the IP polarization is intercorrelated with the OP polarization (Figure 1e and 1f). Interestingly, there is no obvious changes in the corresponding IP phase and IP amplitude images for the 90 nm thick sample after similar switching experiment, as shown in Figure 1i and 1j, while the OP ferroelectricity remains reversible (Figure 1g and 1h). This indicates there may be a critical thickness below which the IP ferroelectricity disappears.

In order to distinguish the critical thickness for the disappearance of IP, both OP and IP PFM scanning were performed to investigate the domain distribution of CIPS flakes with various thicknesses ranging from 320 nm to 30 nm. Figure 2a-h shows both PFM OP and IP phase mappings overlaid on three-dimensional (3D) topography. The height profile of CIPS flakes with different thicknesses can be found in Supporting Information **Figure S4**. For the flakes with thickness 320 nm (Figure 2a-b) and 100 nm (Figure 2c-d), both IP and OP phase images show contrast, indicating that the exfoliated samples have both IP and OP components of ferroelectric polarization. Moreover, we performed the angular-dependent PFM by rotating the samples by 45 and 90 degrees on an identical flake with thickness around 240-250 nm (**Figure S5**), supporting the existence of both IP and OP polarization. When the thickness is reaching 80 nm (Figure 2e-f), OP ferroelectric domains still can be clearly distinguished, as shown in Figure 2e. However, there is no obvious contrast of IP phase image in Figure 2f. The comparison between Figure 2e and 2f clearly demonstrates the existence of OP PFM signal and no hint of IP signal, suggesting the absence of IP ferroelectric domain below 90 nm. Furthermore, combining the thickness information from both Figure 1 and Figure 2, we can conclude that the critical thickness for the disappearance of IP component of ferroelectricity is between 90 and 100 nm. As we thinned down the sample to 30 nm (Figure 2g-h), no phase contrast difference is observed in both OP and IP orientations. More details on various OP and IP PFM images in thicknesses with 250 and 50 nm can be found in Supporting Information **Figure S6**.

To identify that formation of single ferroelectric domain or disappearance of OP polarization below the critical thickness (90 nm), the vertical and lateral phase-voltage hysteresis loops (Figure 2i-l) and amplitude-voltage butterfly loops (**Figure S7**) were carried out for CIPS flakes with different thicknesses. For the 320 nm (Figure 2i and S7a) and 100 nm (Figure 2j and S7b) thick flakes, the butterfly loops of the PFM amplitude signals and the distinct switching of the phase signals further demonstrate the existence of both OP and IP ferroelectric polarizations above 90 nm. When the thickness below 90 nm, especially for the CIPS flakes with thicknesses of 80 and 30 nm, though there is no phase contrast, the vertical piezoresponse (Figure 2k-l and S7c-d) indicates that it is in single OP ferroelectric domain state. However, the comparison between Figure 2i-j and 2k-l, distinct switching of the phase signals in the former and no hint of the lateral piezoresponse in the latter, clearly corroborates the absence of IP ferroelectric domain below 90 nm. Furethermore, the switching spectroscopic loops for 50 nm thick flake, as displayed in Supporting Information **Figure S8**. The distinct switching of the phase signal is also observed, whereas the absence of lateral piezoresponse, further demonstrates that the single ferroelectric domain in OP direction is formed and the disappear of IP ferroelectric polarization below 90 nm.

We also performed the contact resonance force microscope (CRFM) measurement (see Experimental Section) on exfoliated samples with various thicknesses both beyond and below the critical thickness. In the measurement, the contact resonant frequency of the tip-sample system is proportional to the elastic property of CIPS flakes, which is monitored in dual amplitude resonance tracking (DART) mode. The larger the elastic stiffness of the sample is, the stronger the contact resonance frequency exhibits, and vice versa.[25,26] **Figure 3**a shows topographic image of the 80 and 260 nm thick CIPS flakes on silica substrate, the corresponding contact resonance frequency mapping is shown in Figure 3b, and histogram of the CRFM image is plotted in Figure 3c. Note that three peaks suggesting 3 different regions have different contract resonant frequencies. The contact resonant frequency difference ($\Delta f = f_{substrate} - f_{CIPS}$) between the 80 nm thick CIPS flake and the silica substrate is about $\Delta f_1 \sim 3.1$ kHz, while the difference is about $\Delta f_2 \sim 4.8$ kHz for the 260 nm thick flake. It can be seen that 80 nm thick flake possesses higher frequencies compared with 260 nm thick sample. Furthermore, we performed more CRFM measurements on plenty of CIPS flakes with different thicknesses, and the contact resonance frequency difference Δf between the flakes and silica substrate is plotted in Figure 3d. It is interesting that Δf has obvious discontinuity at the thickness about 100 nm, which is in consistent with the critical thickness we obtained from PFM measurement. The elastic properties of various thickness of CIPS further demonstrates that there is a critical thickness between 90-100 nm, below which CIPS flakes possess higher frequencies, meaning that the elastic stiffness becomes larger. Inversely, the flakes are softer above the critical thickness.

The coincidence between PFM and CRFM results suggests a possible phase transition. Therefore, we performed the DFT calculations (see more details in Experimental Section) to understand the disappearance of IP polarization and evolution of elastic property. It is well established that the OP polarization in CIPS contributes to the Cu off-centering displacement.[27-29] As illustrated in **Figure 4**a, the Cu atoms move toward the sulphur plane in ferroelectric structure (the monoclinic structure with space group *Cc*). This Cu off-centering movement not only induces OP displacement (1.32 Å) corresponding to OP polarization (3.14 $\mu C/cm^2$), but also IP displacement (0.18 Å) corresponding to IP polarization (4.48 $\mu C/cm^2$) which is even larger than OP polarization. In addition, it is well known that metal thio- and selenophosphates with chemical formula $M^{1+}M^{3+}[P_2S(Se)_6]^{4-}$ have many structural configurations, among which the trigonal ($P\bar{3}1c$) and centrosymmetric structure has been reported at room temperature for $CuInP_2Se_6$.[24] In order to explore the possible stable phases below the critical thickness of $CuInP_2S_6$ in our experiment, all subgroups of structures between the $P\bar{3}1c$ and *Cc* space groups have been checked here.[30] With symmetry analysis, we found that only the polar structure with *P*31*c* space group satisfies the condition that only OP polarization exists. Therefore, the structure below the critical thickness may belong to the *P*31*c* space group. In order to verify our speculation for thin flakes, two bilayer slab models were built for the *P*31*c* and *Cc* space group to simulate the thinnest CIPS flakes. The calculated results demonstrated that the *P*31*c* bilayers has lower energy than *Cc* bilayer structure by 10 meV, which indicates the *P*31*c* structure could be stabilized compared with *Cc* structure below the critical thickness. Moreover, the Cu off-centering displacement (1.31 Å) in the *P*31*c* model induces OP polarization of 4.22 µC/cm$^2$ (without IP polarization), which is also consistent with our experimental observation. Meanwhile, in comparison to the monoclinic phase, the trigonal phase possesses the shorter van der Waals interlayer distance (3.116 Å in trigonal state *vs* 3.244 Å in monoclinic state) and shorter lattice along the direction perpendicular to *ab* plane (12.995 Å in trigonal state *vs* 13.149 Å in monoclinic state), therefore resulting in higher stiffness than that of monoclinic phase.

We then performed atomic-resolution STEM and electron diffraction measurement to verify this structural phase transition from monoclinic to trigonal phase driven by the thickness reduction, as shown in **Figure 5**. The monoclinic phase structure of bulk CIPS have been demonstrated previously, [27-29,31] as shown in Figure 5a. We performed STEM measurement to identify the CIPS flake structures below 90 nm thickness by using two DFT calculated crystal structures with *Cc* (monoclinic) phase (Figure 5a) and *P*31*c* (trigonal) phase (Figure 5c). The simulated electron diffraction images for *Cc* phase and *P*31*c* phase in plane are shown in Figure 5b and 5d, respectively. Note that the significantly different diffraction patterns for the monoclinic and trigonal structure, clearly indicates the simulated electron diffraction image of *P*31*c* phase is in excellent agreement with that of measurement (Figure 5e). Figure 5f shows that the atomic-resolution STEM image of CIPS along the [001] zone axis. The comparison between STEM measurement with simulation (Figure S9), which also in consistent with our statement that the CIPS transfers to *P*31*c* phase below critical thickness, which results in the disappearance of IP polarization. This result shows that the thin CIPS flakes indeed transfer to *P*31*c* phase when below critical thickness, as our experimental measurements and DFT simulations predict.

The discovery of IP polarization and its rotation in new phase in thin CIPS may induce the morphotropic phase boundary (MPB) in layered materials and enhance the piezoelectricity. As is well known, different phases coexist and easy polarization rotation at MPB are critical for the piezoelectric enhancement.[32,33] However, this polarization rotation at MPB phenomena in layered ferroelectric materials are rare, due to the polarization rotation restricted by the layered structure. And hence it is very limited to optimize the piezoelectric properties of layered materials. Interestingly, with S substituted by Se in layered CIPS, *i.e.* $CuInP_2Se_6$, shows ferroelectric properties (*P*31*c* phase) below 235 K.[34] The solid solution $CuInP_2(S_xSe_{1-x})_6$ transfers from trigonal to monoclinic phases with changing S concentration from *x*=0 to 1.[35,36] Therefore, it is expected that there may coexist of trigonal and monoclinic structures with proper S doping concentration or specific thickness, and the hint of the exsistance of MPB is suggested by carefully analyzing the dielectric properties of $CuInP_2(S_xSe_{1-x})_6$ system.[34] In this case, the polarization rotation may be the reason for such MPB, just like in PZN-PT at MPB,[32] which improves the piezoelectric properties. Such two phases coexist and polarization rotations in layered ferroelectric materials may open a new path for designing and optimizing the piezoelectricity in layered materials, such as Bi-based layered piezoelectrics with high Curie temperature, which are of critical importance in piezoelectric applications at high temperature.[37]

## 3. Conclusion

In conclusion, we revealed that a monoclinic (*Cc* phase) to trigonal (*P*31*c* phase) structural phase transition occurs in vdW ferroelectric CIPS system as the thickness reaching below 90 nm. Such a structural phase transition is suggested by our DFT calculation and verified by TEM measurements, which is also responsible for the abrupt change of elastic properties and the disappearance of IP ferroelectricity in thin CIPS flakes, visualized by PFM technique. The coexistence of IP and OP polarizations in monoclinic phase and polarization rotates to OP in trigonal phase may open a new door to optimize the piezoelectricity in layered materials.

Furthermore, structural change at room temperature with a significant IP non-volatile polarization in vdW ferroelectricity also provides us with a promising framework to study the coupling of the lattice degree of freedom with other order parameters in future micro/nano applications.

**Experimental Section**

**Sample preparation and characterization**

High-quality single crystals of CIPS were synthesized by chemical transportation reaction using the elements in the stoichiometric proportions. The crystals formed as thin and flexible platelets, with irregular shapes and lateral dimensions up to about 1 cm. The single-crystal XRD pattern was collected from a Bruker D8 Advance x-ray diffractometer using Cu $K_\alpha$ radiation. The thin flakes were obtained by mechanical exfoliation from synthetic bulk crystals onto heavily doped silicon substrates. The thickness of the flakes was identified from their optical contrast and atomic force microscopy (AFM).

**Transmission electron microscopy (TEM) measurement**

Note that for TEM experiment, it is difficult to get atomic resolution for thicker samples, usually above 100 nm. The samples were prepared by dropping the alcohol dispersion onto the carbon-coated copper TEM grids using pipettes and dried under ambient condition. A piece of sample was put into some alcohol and then we applied ultrasonic treatment for ~ 3 min. As a result, the flake thickness was far below the critical thickness. TEM and SAED images were taken from Tecnai F20 at 200 kV. The atomically resolved STEM images was carried out on an aberration-corrected FEI Titan Themis G2 microscope operate at 300kV with a beam current of 45 pA. STEM simulation was carried out by Kirkland with computer software.[38] The STEM simulated parameters were set with the beam energy of 300 keV (accelerating voltage), object aperture of 25.1 mrad (convergence semiangle), and a STEM ADF detector of 53-260 mrad (collection semi-angle). Besides, the transmission function size was 2048 × 2048 pixels and STEM probe size was 512 × 512 pixels. The other parameters were set as default.

**Piezoresponse force microscopy (PFM) measurement**

PFM measurement was performed using a commercial atomic force microscope (Asylum Research MFP-3D) with Ti/Ir-coated Si cantilever tips and diamond-coated Si cantilever tips, respectively. In resonance-enhanced mode, a soft tip with a spring constant of ~2.8 N m$^{-1}$ was driven with an ac voltage ($V_{ac}$ = 0.5-1 V) under the tip-sample contact resonant frequency (~320 kHz). The OP and IP PFM are acquired at the drive frequency of ~320 kHz and ~730 kHz in Vector PFM mode, respectively. The vertical and lateral piezoresponse hysteresis loops were measured by dual ac resonance tracking PFM (DART-PFM) mode in CIPS with various thicknesses. The local piezoresponse hysteresis loops were measured at three different arbitrary points for three times at each position.

**Contact resonance force microscope (CRFM) measurement**

The CRFM measurement was performed using a MFP-3D AFM with AC240TS-R3 cantilever tips. In the measurement, a very low amplitude vertical modulation was applied to the samples by an external actuator. A constant force about 85 nN was applied at the samples while the cantilever scanning along the samples' surface. The contact resonance frequency of the cantilever in contact with the surface monitored in the DART mode strongly depends on the coupling with the mechanical properties of the samples. The relationship between the normalized contact stiffness $k^*/k_{lever}$ and contact resonant frequency $f_n$ can be expressed as follows:[39-40]

$$\frac{k^*}{k_{lever}} = \frac{2}{3}(x_n L \gamma)^3 \frac{(1 + cosx_n L coshx_n L)}{B}, (where\ x_n L = x_n^0 L \sqrt{f_n / f_n^0}) \qquad (1)$$

The $k^*/k_{lever}$ as a function of the relative tip position $\gamma$ for each mode $n$. Since $k^*/k_{lever}$ should be the same for the first and second modes at one physical value of $\gamma$, it can be determined along with $\gamma$ from where the two modes intersect each other. Thus, contact stiffness can be described by contact resonant frequency $f_n$ and relative tip position $\gamma$, as shown in Equation (1). Therefore, as the stiffness of the samples contact changes, the frequency of the contact resonance changes: the higher stiffness will induce the higher frequency.[41]

**Density Functional Theory (DFT) Calculation**

All calculations in this work was performed using the Vienna Ab initio Simulation Package (VASP)[42,43] based on DFT. Perdew-Burke-Ernzerhof functional of generalized gradient approximation was chosen for the exchange-correlation functional. In the calculation, a kinetic energy cutoff of 500 eV for the plane wave expansion and 6 · 3 · 1 grid of k points were set for the bilayer CIPS. For the bilayer models, a vacuum layer of 20 Å was used to avoid the image interaction due to the periodic boundary condition. The energy and force convergence criterions are $10^{-6}$ eV and $10^{-3}$ eV/Å$^{-1}$ in the relaxation, respectively. In order to take into account the interlayer van der Waals interaction in bulk CIPS, DFT-D2 method was used here. Additionally, the Berry-phase method was employed to calculate the polarization. The lattice parameters of bulk CIPS could be found in **Table S1** and **Figure S10** in supporting information.

**Acknowledgements**
The authors are grateful for the fruitful discussions with Lu You from Nanyang Technological University and Prof. Sang-Wook Cheong from Rutgers University. The work at Beijing Institute of Technology is supported by National Natural Science Foundation of China with Grant Nos. 11572040, 11604011, 11804023, and the China Postdoctoral Science Foundation with Grant No. 2018M641205. The work at Peking University is supported by National Natural Science Foundation of China (grant 51672007). Theoretical calculations were performed using resources of the National Supercomputer Centre in Guangzhou, which was supported by Special Program for Applied Research on Super Computation of the NSFC-Guangdong Joint Fund (the second phase) under Grant No. U1501501. We gratefully acknowledge Electron Microscopy Laboratory in Peking University for the use of Cs corrected electron microscope.

**Author contributions**
X.W. and J.H. designed and supervised the experiment and theoretical calculations. J.D. and X.W. prepared the samples, performed PFM experiments. J.D. and Y.L. carried out elastic property characterization and data analysis. Y.L. P.L. and J.H. performed the theoretical calculation. M.L. and P.G. carried out the TEM characterization. S.X. and T.X. performed X-ray single crystal diffraction. J.D., Y.L. J.H. and X.W. co-wrote the paper. All authors discussed the results and commented on the manuscript.

**Conflict of Interest**

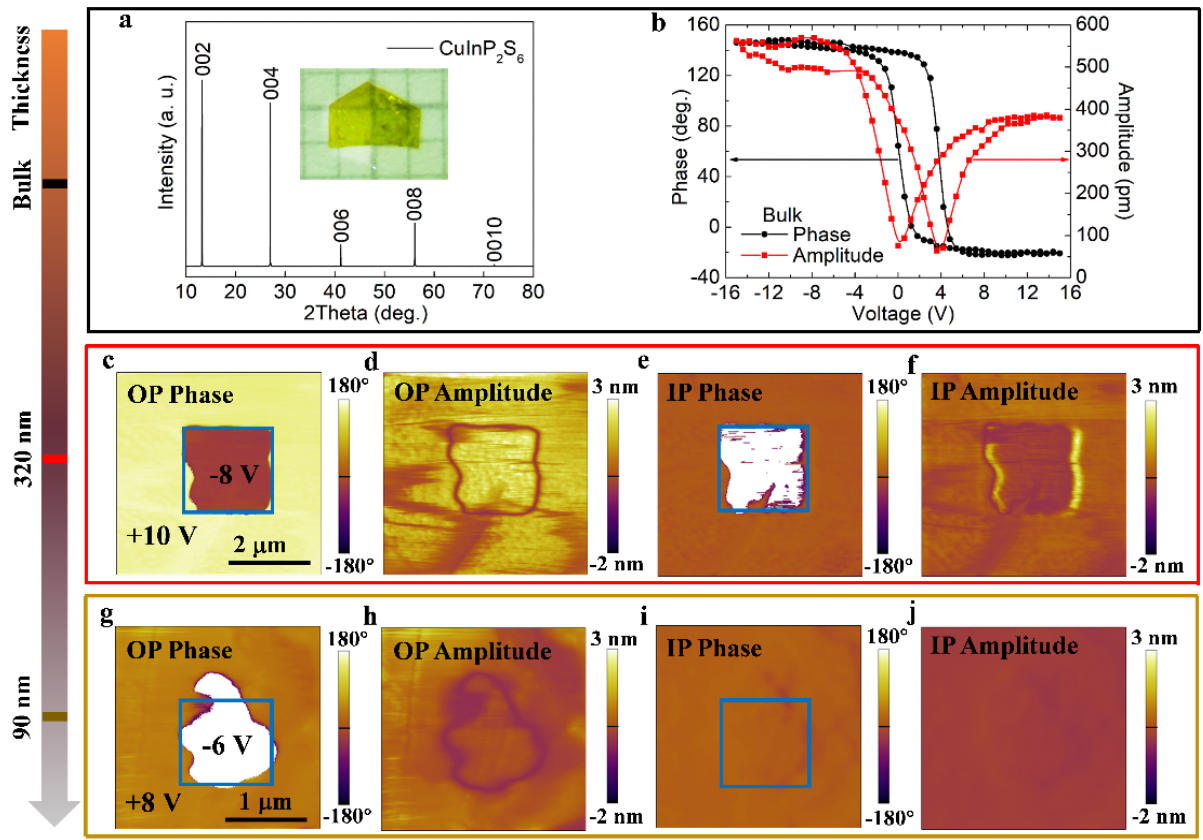

**Figure 1.** Characterization and ferroelectric switching of bulk and exfoliated CIPS with thickness of 320 and 90 nm. (a) The X-ray diffraction for CIPS and the inset is single-crystal specimen placed on a millimeter grid. (b) Switching Spectroscopy PFM for bulk single crystals. (c) OP phase, (d) OP amplitude and the corresponding (e) IP phase, (f) IP amplitude images of a 320 nm thick CIPS flake acquired immediately after writing one square pattern centrically by applying -8 and +10 V voltages consecutively. The scale bar is 2 μm. (g) OP phase, (h) OP amplitude and the corresponding (i) IP phase, (j) IP amplitude images of a 90 nm thick CIPS flake acquired immediately after writing one square pattern centrically by applying -6 and +8 V voltages consecutively. The scale bar is 1 μm.

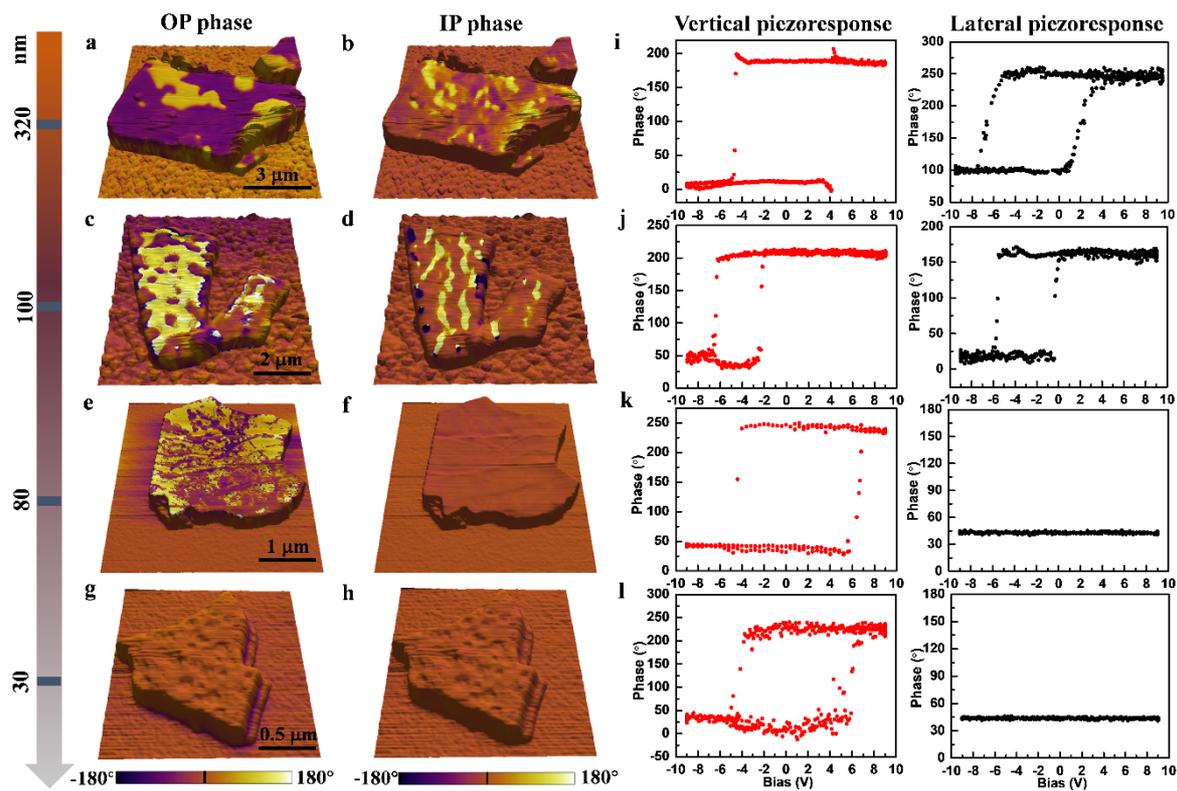

**Figure 2.** Piezoresponse force microscopy and switching spectroscopy loops of CIPS flakes with thickness ranging from 320 to 30 nm. (a-h) The OP and IP phase imaging overlaid on 3D topography, respectively. (i-l) Vertical and lateral phase-voltage hysteresis loops obtained from 320, 100, 80, 30 nm thick CIPS flakes, respectively. Red dots for vertical and black dots for lateral signals.

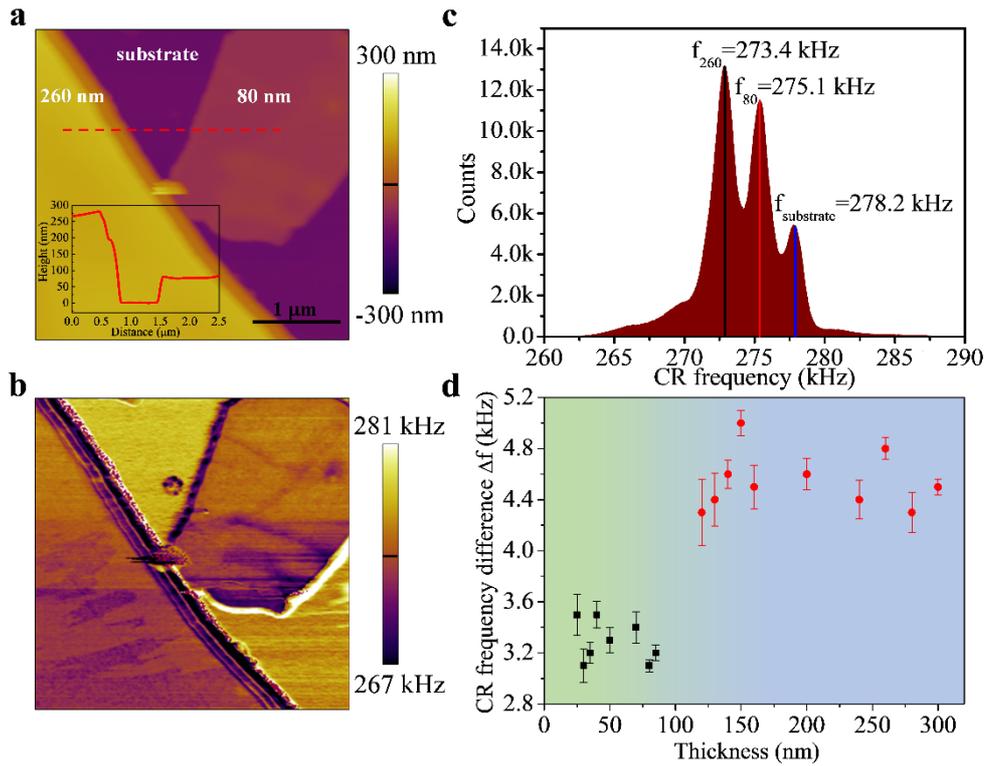

**Figure 3.** Elastic property for CIPS flakes with various thicknesses. (a) Topography of samples. The thickness values are labeled in the corresponding flakes and the inset is height profile. The scale bar is 1 μm. (b) Contact resonance (CR) frequency mapping of the CIPS flakes and silica substrate. (c) Histogram of the contact resonance force microscope (CRFM) image shown in (b): black, red and blue lines correspond to CR frequency of 260 nm thick CIPS flake, 80 nm thick sample and silica substrate, respectively. (d) The CR frequency difference between the flakes and silica substrate ($\Delta f = f_{substrate} - f_{CIPS}$) for CIPS flakes with various thicknesses, showing abrupt change of elastic properties of flakes below 90 nm.

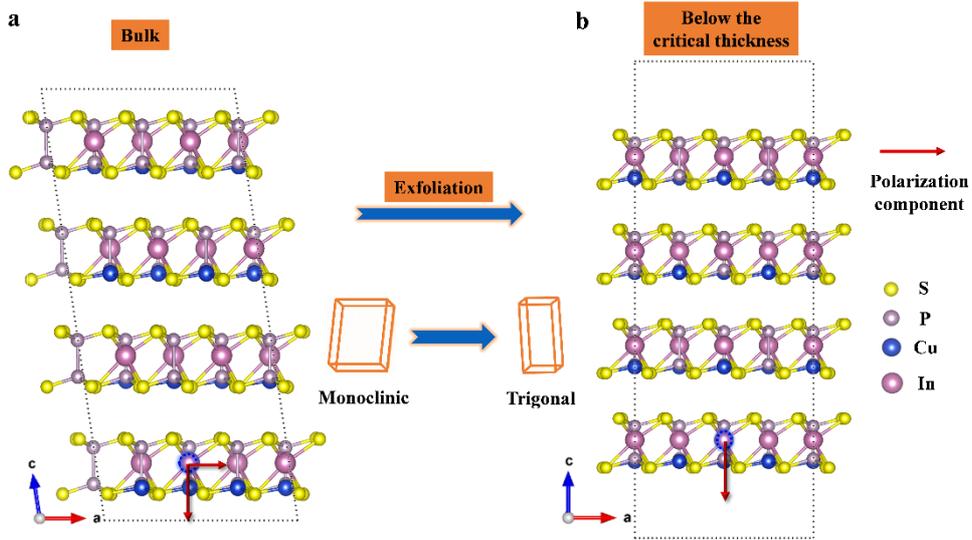

**Figure 4.** Schematics for the origin of OP and IP polarization of the bulk CIPS and disappearance of IP polarization in CIPS below the critical thickness. (a) The monoclinic structure in bulk ferroelectric CIPS. (b) The CIPS below the critical thickness of trigonal structure. The insets show schematics for monoclinic and trigonal structure. The red arrows indicate polarization component, resulting from the displacement of Cu atoms relative to its position (blue doted circles) in the corresponding centrosymmetric structures.

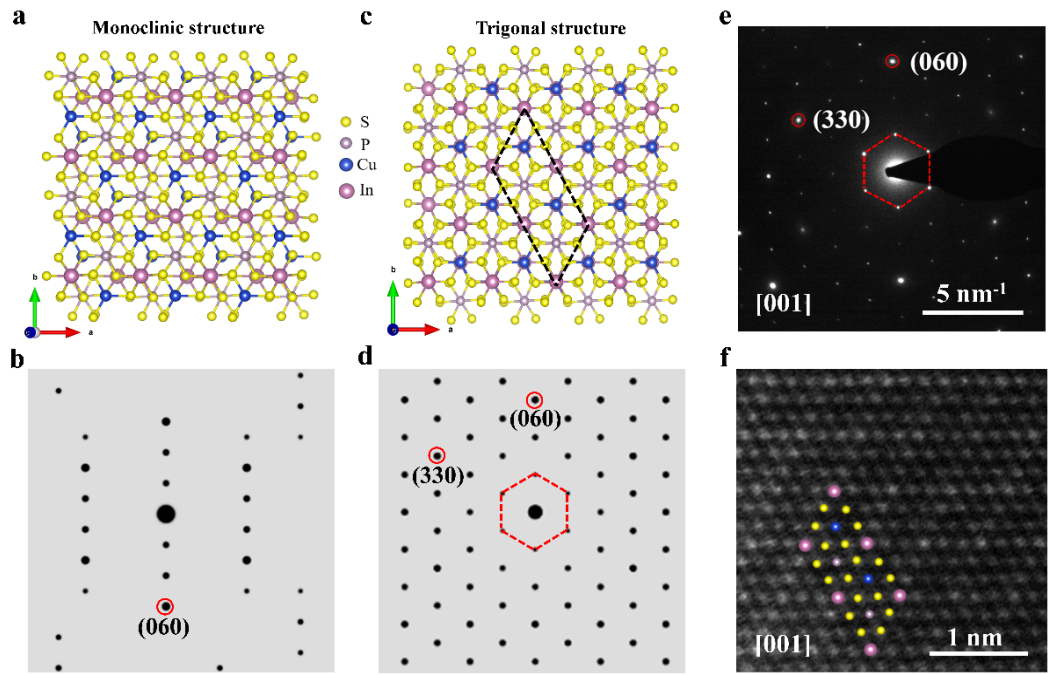

**Figure 5.** Atomic structure characterization of CIPS below the critical thickness. (a) Monoclinic (*Cc*) structure of CIPS viewed along $c^*$ direction (out-of-plane direction). (b) The simulated electron diffraction image of monoclinic structure CIPS along $c^*$ direction. (c) Trigonal (*P*31*c*) structure of CIPS viewed along [001] axis. (d) The simulated electron diffraction image of trigonal structure CIPS along [001] axis. (e) Electron diffraction image of thin CIPS along the the [001] axis. (f) Atomic-resolution STEM image of thin CIPS seen along the [001] axis.